\documentclass[9pt]{article}
\usepackage{extsizes}
\usepackage[super,sort&compress,comma]{natbib} 
\usepackage[version=3]{mhchem}
\usepackage[left=1.5cm, right=1.5cm, top=1.785cm, bottom=2.0cm]{geometry}
\usepackage{balance}
\usepackage{mathptmx}
\usepackage{sectsty}
\usepackage{graphicx} 
\usepackage{lastpage}
\usepackage[format=plain,justification=justified,singlelinecheck=false,font={stretch=1.125,small,sf},labelfont=bf,labelsep=space]{caption}
\usepackage{float}
\usepackage{fancyhdr}
\usepackage{fnpos}
\usepackage[english]{babel}
\addto{\captionsenglish}{%
  
}
\usepackage{array}
\usepackage{droidsans}
\usepackage{charter}
\usepackage[T1]{fontenc}
\usepackage[usenames,dvipsnames]{xcolor}
\usepackage{setspace}
\usepackage[compact]{titlesec}
\usepackage{hyperref}
\usepackage{xcolor}
\usepackage{epstopdf}

\definecolor{cream}{RGB}{222,217,201}

\begin{document}

\pagestyle{fancy}
\thispagestyle{plain}
\fancypagestyle{plain}{
\renewcommand{\headrulewidth}{0pt}
}

\makeFNbottom
\makeatletter
\renewcommand\LARGE{\@setfontsize\LARGE{15pt}{17}}
\renewcommand\Large{\@setfontsize\Large{12pt}{14}}
\renewcommand\large{\@setfontsize\large{10pt}{12}}
\renewcommand\footnotesize{\@setfontsize\footnotesize{7pt}{10}}
\makeatother

\renewcommand{\thefootnote}{\fnsymbol{footnote}}
\renewcommand\footnoterule{\vspace*{1pt}%
\color{cream}\hrule width 3.5in height 0.4pt \color{black}\vspace*{5pt}} 
\setcounter{secnumdepth}{5}

\makeatletter 
\renewcommand\@biblabel[1]{#1}            
\renewcommand\@makefntext[1]%
{\noindent\makebox[0pt][r]{\@thefnmark\,}#1}
\makeatother 
\renewcommand{\figurename}{\small{Fig.}~}
\sectionfont{\sffamily\Large}
\subsectionfont{\normalsize}
\subsubsectionfont{\bf}
\setstretch{1.125} 
\setlength{\skip\footins}{0.8cm}
\setlength{\footnotesep}{0.25cm}
\setlength{\jot}{10pt}
\titlespacing*{\section}{0pt}{4pt}{4pt}
\titlespacing*{\subsection}{0pt}{15pt}{1pt}

\fancyfoot{}
\fancyhead{}
\renewcommand{\headrulewidth}{0pt} 
\renewcommand{\footrulewidth}{0pt}
\setlength{\arrayrulewidth}{1pt}
\setlength{\columnsep}{6.5mm}
\setlength\bibsep{1pt}

\makeatletter 
\newlength{\figrulesep} 
\setlength{\figrulesep}{0.5\textfloatsep} 

\newcommand{\topfigrule}{\vspace*{-1pt}%
\noindent{\color{cream}\rule[-\figrulesep]{\columnwidth}{1.5pt}} }

\newcommand{\botfigrule}{\vspace*{-2pt}%
\noindent{\color{cream}\rule[\figrulesep]{\columnwidth}{1.5pt}} }

\newcommand{\dblfigrule}{\vspace*{-1pt}%
\noindent{\color{cream}\rule[-\figrulesep]{\textwidth}{1.5pt}} }

\makeatother

\noindent\LARGE{\textbf{Near-zero surface pressure assembly of rectangular lattices of microgels at fluid interfaces for colloidal lithography}} \\
\vspace{0.3cm} \\

\large{Miguel Angel Fernandez-Rodriguez,\textit{$^{a,b}$$^{\ast}$} Maria-Nefeli Antonopoulou,\textit{$^{b,c}$} and Lucio Isa\textit{$^{b}$} } \\

\noindent\normalsize{Understanding and engineering the self-assembly of soft colloidal particles (microgels) at liquid-liquid interfaces is broadening their use in colloidal lithography. Here, we present a new route to assemble rectangular lattices of microgels at near zero surface pressure relying on the balance between attractive quadrupolar capillary interactions and steric repulsion among the particles at water/oil interfaces. These self-assembled rectangular lattices are obtained for a broad range of particles and, after deposition, can be used as lithography masks to obtain regular arrays of vertically aligned nanowires via wet and dry etching processes.} \\


\renewcommand*\rmdefault{bch}\normalfont\upshape
\rmfamily


\footnotetext{\textit{$^{a}$}~\textit{Laboratory of Surface and Interface Physics, Biocolloids and Fluid Physics group, Department of Applied Physics, Faculty of Sciences, University of Granada, Campus de Fuentenueva s/n, ES 18071 Granada, Spain. E-mail: mafernandez@ugr.es}}
\footnotetext{\textit{$^{b}$}~\textit{Laboratory for Soft Materials and Interfaces, Department of Materials, Swiss Federal Institute of Technology Z\"urich, Vladimir-Prelog-Weg 1-5/10, 8093 Z\"urich, Switzerland.}}
\footnotetext{\textit{$^{c}$}~\textit{Current address: Polymeric Materials, Department of Materials, Swiss Federal Institute of Technology Z\"urich, Vladimir-Prelog-Weg 1-5/10, 8093 Z\"urich, Switzerland.}}


Microgels are versatile soft colloids \cite{Lyon-Nieves} that self-assemble at water/air and water/oil interfaces \cite{Karg_Richtering2019,Rey2020} enabling their controlled deposition on a solid substrate and their use for soft colloidal lithography \cite{Rey_NanoLett_2016,Fernandez-Rodriguez_Nanoscale_2018}. Although microgels made of Poly(N-isopropylacrylamide) (PNIPAM) are hydrophilic and easily dispersed in water, they can be readily adsorbed at liquid-liquid interfaces using a spreading agent. Upon adsorption at the interface, each microgel spreads to reduce the area of the bare fluid-fluid interface until the deformation is counterbalanced by the microgel internal elasticity \cite{Style2015,Mehrabian2016}.  The spreading can be tuned by adjusting the particle crosslinking density, leading to effective diameters at the interface greater than twice the diameter in bulk dispersions \cite{Camerin_ACSNano_2019}. Once at the interface, the structural and dynamical behaviour of assemblies of microgels is mainly dictated by capillary and steric interactions \cite{Huang2016}. Upon compression, microgel monolayers can reach a point at which all microgels enter into contact to form hexagonal lattices whose lattice spacing can be reduced by further compression \cite{Rey_SoftMatter_2016, Fernandez-Rodriguez_Nanoscale_2018}. In order to expand the applications of soft colloidal lithography, new routes need to be developed to produce patterns with symmetries beyond the hexagonal one.  
A broad range of Bravais lattices can be obtained for particles interacting via soft electrostatic repulsion via the stretching of hexagonal lattices while they are being deposited on the substrate \cite{Hummel_Langmuir_2019}. However, this process relies on a delicate interplay between the wettability of the substrates, the angle of deposition and a thermal fixation of the colloids to circumvent the collapse of the structures due to capillary forces during drying. Additional structures, including binary colloidal alloys \cite{Fernandez-Rodriguez_Nanoscale_2018} and Moiré and complex tessellations \cite{Volk_PCCP_2019,Nature_2D}, can also be obtained by sequential depositions of microgels, respectively of two different sizes or of the same size. Nevertheless, they require multiple steps and a careful control of the surface pressure $\Pi$ to achieve the target structure. In order to circumvent sensitive processes, it would be desirable to obtain directly non-hexagonal structures via robust self-assembly at the fluid interface. The existence of complex two-dimensional structures beyond the triangular symmetry was numerically predicted by Jagla twenty years ago \cite{Jagla_JChemPhys_1999} for colloids with a repulsion potential exhibiting two distinct length scales (e.g. a hard core and a soft shell) and recently experimentally realised by Rey et al. for polystyrene colloids with a microgel soft shell confined at a water/oil interface \cite{Rey_JACS_2017}. Non-triangular assemblies, including strings and rectangular lattices, occur at high values of $\Pi$, beyond compact hexagonal monolayers, where the microgel shells are under compression. However, the parameter space in the phase diagram where such structures are obtained is fairly small, and their kinetic access is delicate, making their experimental realization a complex task with limited tuneability. 

We report here a novel, robust route to obtain self-assembled rectangular microgels lattices at fluid interfaces for a broad range of different particles. In contrast to the lattices predicted by Jagla \cite{Jagla_JChemPhys_1999}, we find rectangular lattices at $\Pi\approx0\,mN/m$, arising from the combination of steric repulsion and leading-order quadrupolar capillary attraction between microgels at the interface, which are  the result of the nanometre-scale roughness induced by the microgels adsorbed at the interface \cite{Danov_JCIS_2005,Stebe_review}. 

\begin{figure}[H]
\centering
  \includegraphics[width=\columnwidth]{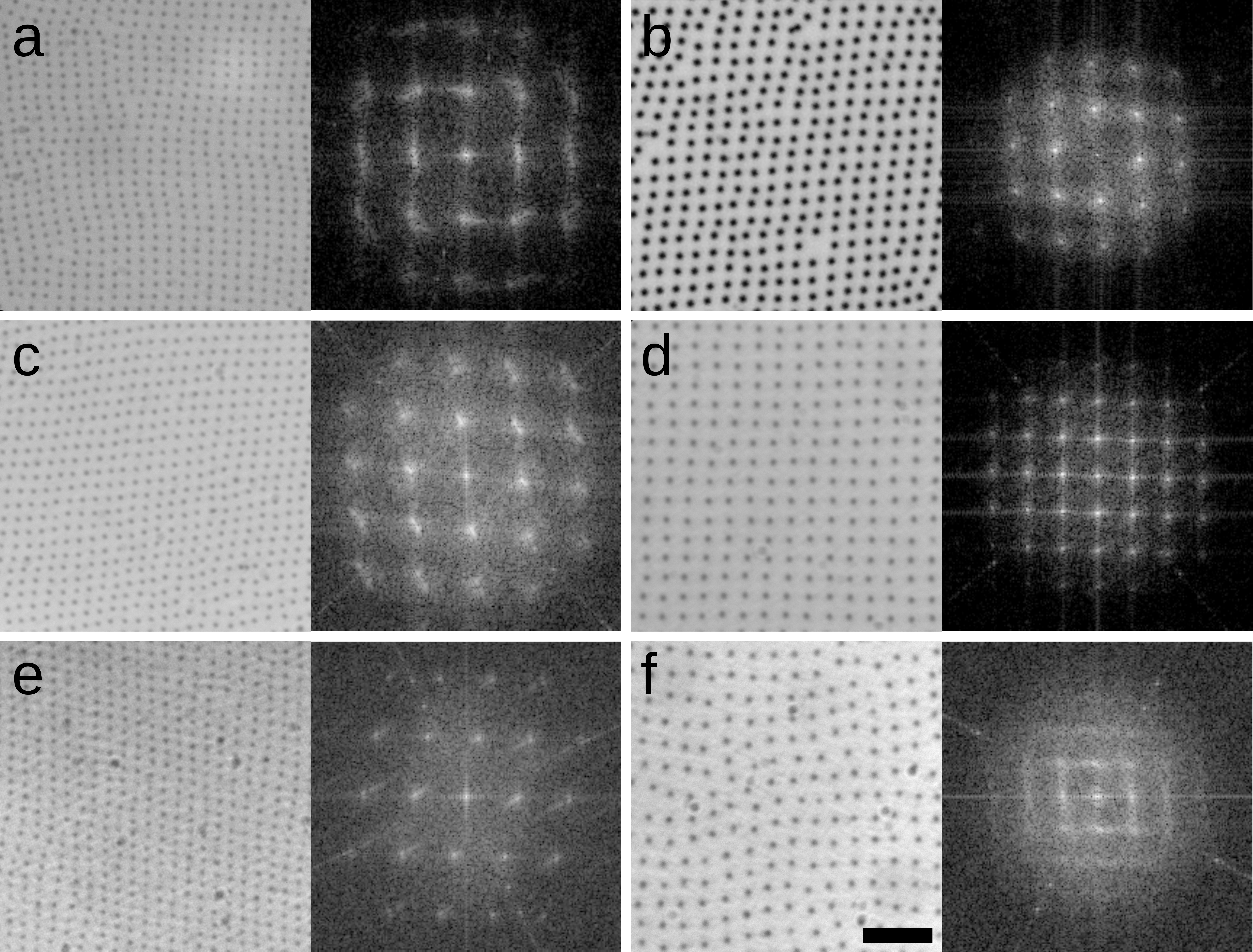}
  \let\nobreakspace\relax\caption{Square lattices and corresponding FFT, obtained by depositing the different microgels \textbf{(a)} C7S0, \textbf{(b)} C7S1, \textbf{(c)} C5S0, \textbf{(d)} C5S1, \textbf{(e)} C3S0, and \textbf{(f)} C3S1, at $\Pi\approx0\,mN/m$. Scale bar is $5\,\mu m$ and the FFTs are enlarged 3 times.}
  \label{fig:0}
\end{figure}

The microgels used here were synthesized via precipitation polimerization of N-Isopropylacrylamide (NIPAM) in water \cite{Camerin_ACSNano_2019}. The NIPAM monomer and crosslinker (BIS) were added in one or two steps to grow larger shells as reported in a previous work \cite{Nature_2D}. Their labeling is CXSY, where X = 3,5,7 \% is the mass ratio between crosslinker and total monomer, whereas the total amount of monomer is constant, and Y=0,1 is the number of steps for additional shell growth. This procedure leads to the synthesis of soft colloidal particles with consistent chemical composition and core-shell architecture, but varying degrees of crosslinking densities and diameters in bulk aqueous suspensions and at the interface. Our microgels have bulk hydrodynamic diameters ranging from $(574\pm73)\,nm$ of C7S0 to $(879\pm121)\,nm$ of C3S1, as measured by DLS (Malvern Zetasizer, z-averaged diameter) and interfacial sizes ranging from $(786\pm30)\,nm$ of C7S0 to $(1578\pm46)\,nm$ of C3S1 (measured ex-situ after deposition on a silicon substrate via AFM, see ESI for full synthesis details and dimensions). In Fig. \ref{fig:0}, we show rectangular lattices obtained after the deposition of monolayers formed at water/hexane interfaces on silicon substrates using a Langmuir-Blodgett (LB) trough at $\Pi\approx0\,mN/m$ for all types of microgel studied. 

\begin{figure}[H]
\centering
  \includegraphics[width=0.8\columnwidth]{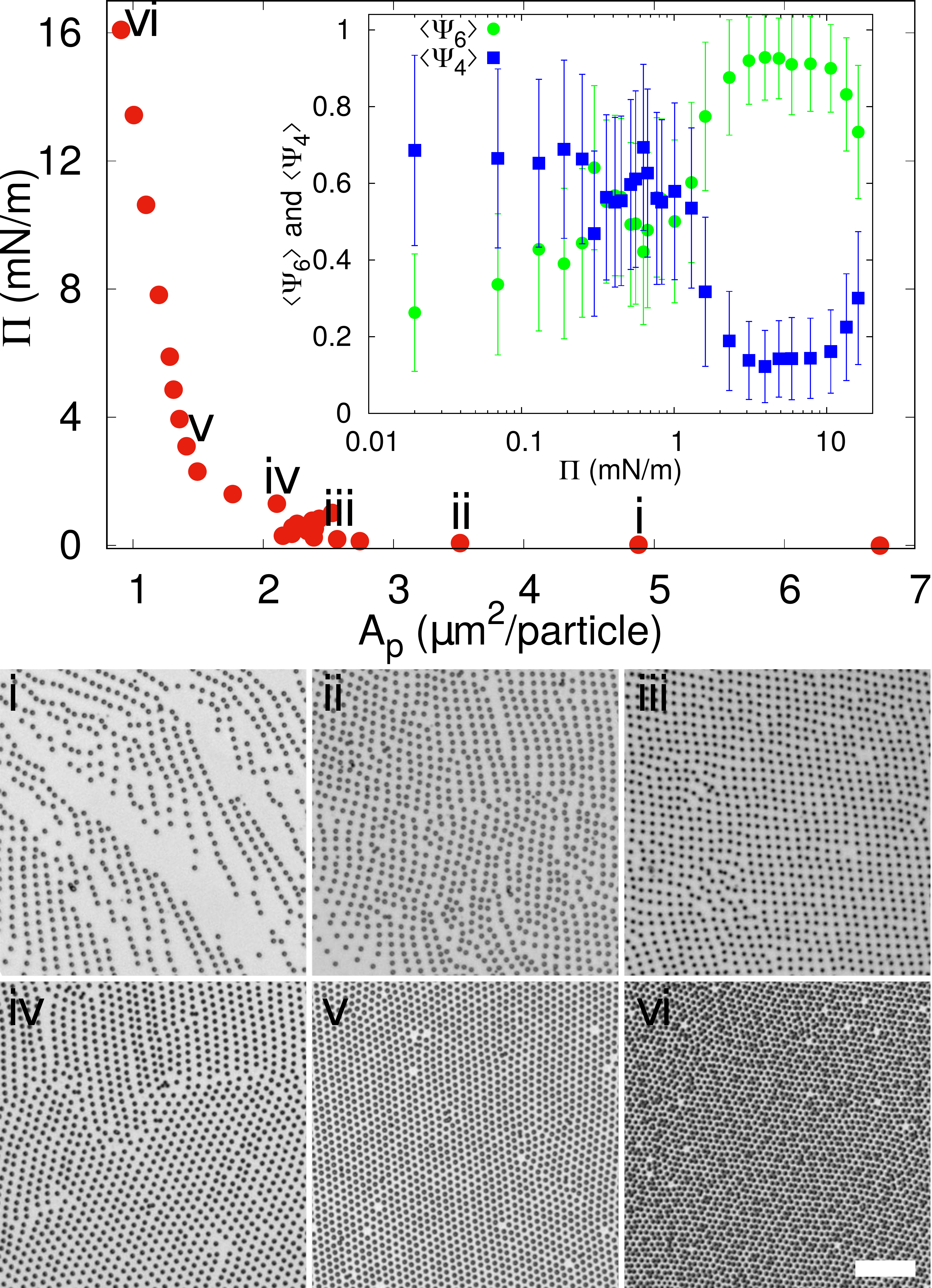}
  \caption{Compression curve for C7S1 microgels as they are deposited on a silicon substrate in an LB trough. Microscope images are provided for different points in the compression curve on the substrate (scale bar is $10\,\mu m$). The inset shows the corresponding order parameters $\langle\Psi_6\rangle$ and $\langle\Psi_4\rangle$ vs $\Pi$. Error bars correspond to standard deviations.}
  \label{fig:1}
\end{figure}

In order to understand in which region of a 2D phase diagram the structures are obtained, in Fig. \ref{fig:1} we show the compression curve of the C7S1 microgels and the corresponding monolayer microstructure after deposition. These experiments are performed on a LB trough, where we deposit the particles on a substrate, which crosses the water/hexane interface at the same time as the monolayer is gradually compressed by the moving barriers of the trough. This procedure allows transferring monolayers subjected to continuously varying surface pressure onto a single sample, so that different locations on the sample correspond to different values of pressure and surface coverage (see ESI for full details). From the reading of $\Pi$ using a Wilhemy plate and the direct measurement of the area per particle $A_p$ (extracted by image analysis after deposition), we observe that lattices with a four-fold coordination (e.g. rectangular and rhomboidal) are found for values of $\Pi\approx0\,mN/m$ in the compression curve. As $\Pi$ grows, the structures transition to the well-known hexagonally packed monolayers. The transition between the two types of coordination is clearly displayed in the inset to Fig. \ref{fig:1}, which shows the average order parameters $\langle\Psi_4\rangle$ and $\langle\Psi_6\rangle$ calculated as in Equation 1, where $j=4,6$ represents the symmetry of the coordination described by $\langle\Psi_j\rangle$, $N$ is the total number of particles, the index $k$ refers to a given particle with $N_l$ neighbours and $\theta_{lk}$ is the angle formed between the particle $k$ and its neighbour $l$ respect to a fixed axis.

\begin{equation}
  \langle\psi_j\rangle=\frac{1}{N}\sum_k^N\left\lvert \frac{1}{N_l}\sum_l^{N_l} e^{ij\theta_{lk}}\right\rvert 
\end{equation}

The average $\langle\Psi_4\rangle$ and $\langle\Psi_6\rangle$ parameters are closer to 1 when there is a mainly rectangular or hexagonal lattice, respectively \cite{Nature_2D}. Therefore, from the data it is clear that for $\Pi\approx0\,mN/m$ $\langle\Psi_4\rangle>\langle\Psi_6\rangle$, while the point at which $\Pi$ starts rising corresponds to a transition to an hexagonal layer, where $\langle\Psi_4\rangle<\langle\Psi_6\rangle$. Above 10 mN/m an isostructural phase transition occurs, which results in a decrease of $\langle\Psi_6\rangle$, as previously reported \cite{Rey_SoftMatter_2016}. 

\begin{figure}[h!]
\centering
  \includegraphics[width=\columnwidth]{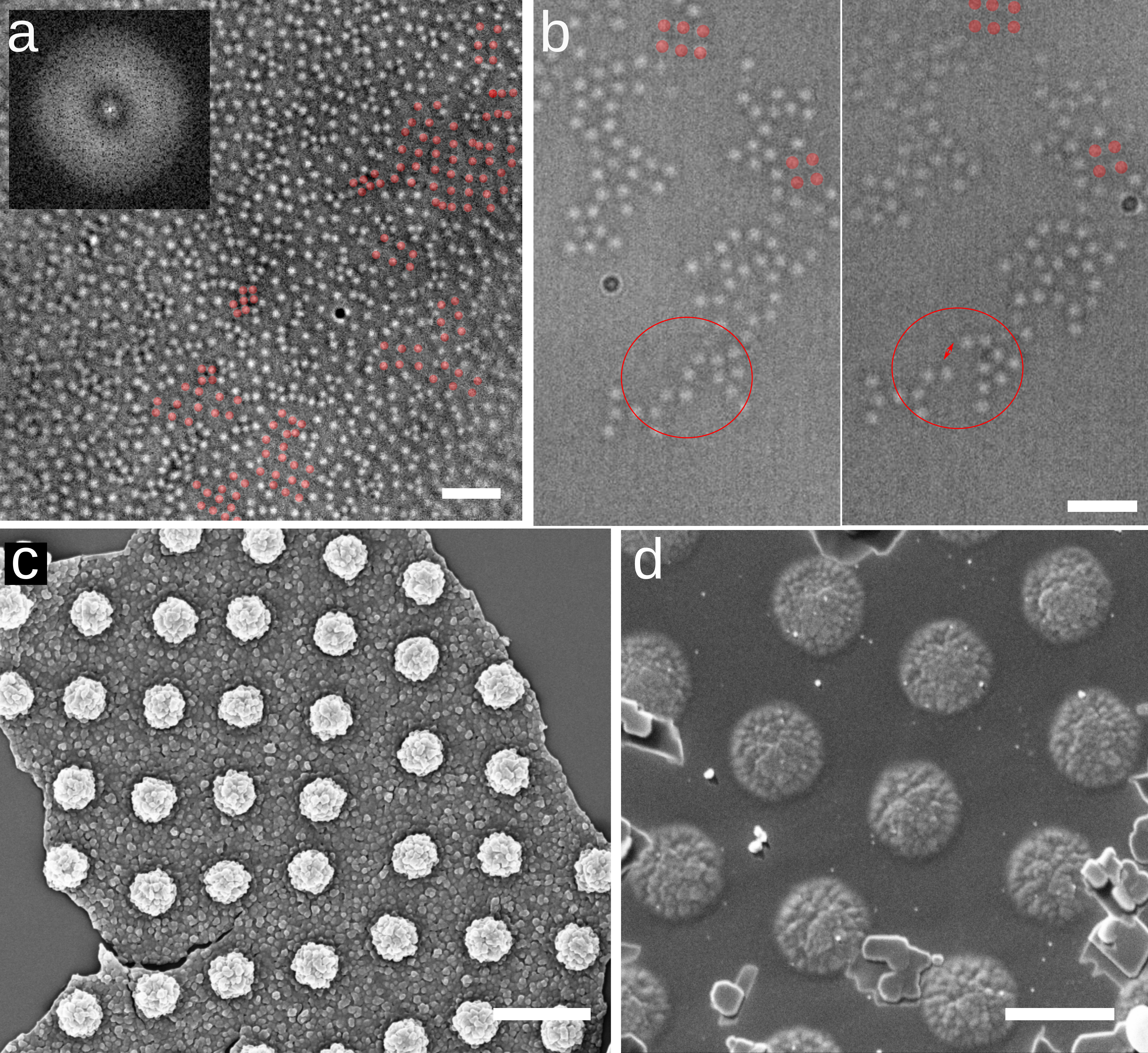}
  \caption{Spontaneous assembly of square lattices of C7S1 microgels at the interface. \textbf{(a)} Frame of Movie S1 at the water/hexadecane interface at the center of a $800\,\mu m$-diameter ring (scale bar: $10\,\mu m$). The inset shows the FFT, with the diffuse ring characteristic of 2D disordered liquids. \textbf{(b)} Two frames of Movie S2 of a water/hexadecane interface showing the spontaneous detachment of a microgel from a cluster, marked for clarity (scale bar: $5\,\mu m$). Representative four-fold coordinated clusters are highlighted in red in (a-b). \textbf{(c)} Cryo-SEM image of a water/decane interface where the interface has been fractured and the oil phase removed, showing the microgels protruding from the vitrified water phase (scale bar: $1\,\mu m$). \textbf{(d)} SEM image of an upside-down fragment of the microgel-cyanoacrylate film, showing the portion immersed in water, with the microgels in a square lattice configuration (scale bar: $2\,\mu m$).}
  \label{fig:2}
\end{figure}

In order to eliminate the possibility that the rectangular lattices are an artifact induced by the deposition process, e.g. as reported for drying droplets \cite{Marin}, we imaged these microgels directly at the water/hexadecane interface, as well as at an water/air interface, in conditions of $\Pi\approx0\,mN/m$ using several complementary approaches. The first approach is direct optical microscopy at the fluid interface using a 40x dipping objective (Nikon) to image the interface through the oil. To reduce the large convection flows present in the system, we corralled the particles using a polymeric ring of $800\,\mu m$-diameter fabricated with a Nanoscribe (fabrication described in ESI) and placed at the microgel-laden water/hexadecane interface with a micropositioner. The field of view is $85\times85\,\mu m^2$ and we imaged close to the center of the ring, avoiding possible confinement effects that are known to provide capillary forces inducing square lattice formation at interfaces \cite{Wurger_PhysRevE_2006,Ershov_PNAS_2013} (see Fig. \ref{fig:2}a). Within the ring, microgels exhibit Brownian motion at the interface (see Movie S1 in ESI) and the overall structure of the interface is globally disordered, as reflected by the FFT in the inset to Fig. \ref{fig:2}a. However, local, clearly four-fold coordinated particle clusters are present at the interface (Fig. \ref{fig:2}a). These clusters are weakly bound, and continuously form and disappear under thermal fluctuations and the residual small drift at the interface (see Fig. \ref{fig:2}b and Movies S1-3). These observations already demonstrate that short-ranged four-fold coordinated assemblies exist at an unperturbed interface. However, the deposition process in an LB trough promotes the creation of larger-area structures. In particular, the presence of the macroscopically straight three-phase (water-oil-silicon wafer) contact line aligns the formed clusters with respect to the deposition direction and the continuous compression of the barriers produces extended regions with the right surface coverage to obtain longer-ranged order. This alignment at the meniscus also explains the formation of lines in Fig. \ref{fig:1}i for lower surface coverages, with insufficient particle numbers to create a continuous monolayer.
As a complementary set of measurements, we employed two different ways of immobilizing and imaging the particle at a fluid interface. Rectangular lattices were first observed in freeze-fractured shadow-cast (FreSCa) cryo-SEM images at the water/decane interface (see Fig. \ref{fig:2}c and ESI), a technique that exploits shock-freezing to immobilize and characterize the microgels in-situ at the water/oil interface \cite{Geisel2012,Camerin_ACSNano_2019}. In a last set of measurements, we placed a silicon wafer in a Teflon beaker filled with water, forming a $\approx30^\circ$ angle with the interface and partially crossing it. We then deposited C7S1 microgels at the water/air interface ensuring that we are close to $\Pi\approx0\,mN/m$. Then, the Teflon beaker was enclosed in a methacrylate box next to a droplet of cyanoacrylate. The evaporating cyanoacrylate reacts at the water/air interface, and after an overnight wait, it created an acrylate film that fixed the particles at the interface\cite{Weiss}. Next, the substrate was extracted from water, collecting the film, which was characterized by SEM and AFM and showed the presence of rectangular lattices (Fig. \ref{fig:2}d and ESI). This experiment indicates that the self-assembly of rectangular lattices is not specific to water/alkane oil interfaces, but it is also found at water/air interfaces, expanding its generality and applicability.

After reporting these observations, we show that the formation of the four-fold coordinated structures can be rationalized by balance between capillary interactions and steric repulsion among the microgels. Although capillary forces are typically associated with interfacial deformations imparted by hard colloids, they have also been reported for soft particles\cite{Rauh_SoftMatter_2017,Huang2016}. In general, the capillary forces between colloidal particles trapped at a fluid interface and displaying an arbitrarily undulated contact line can be described as a multipole expansion, where, in the absence of gravity, the leading-order term has a quadrupolar symmetry \cite{Stamou}. Even if higher-order terms may not be neglected for a full quantification of capillary interactions at close separations \cite{Near_field,Bowden,Stebe_review}, the clear presence of four-fold symmetries in our experiments confirms that the aggregation behavior can be captured by the leading quadrupolar term.

The steric repulsion between microgels at interfaces has been described as via Hertzian or generalized Hertzian potentials \cite{Nature_2D, Camerin2020}. As we are in the regime where the microgels are not externally compressed, the specific form of the steric potential is not relevant and its contribution is effectively 0 until the particles overlap. However, we can consider the steric repulsion, once the particles enter into contact, to be much steeper than the quadrupolar attractive interaction. This implies that, essentially, capillary forces bring the particles at a center-to-center distance approximately equal to their single-particle diameter at the interface, and not closer. In order to reduce this separation further, external compression, i.e. with the barriers of a LB trough, is required and leads to the formation of 2D hexagonal crystals with lattice parameters smaller than the single-particle diameter. The fact that the microgels can rearrange at the interface, i.e. that the clusters spontaneously form and disappear under thermal fluctuations, furthermore indicates that the attractive capillary potential at contact is of the order of a few $K_BT$.
With these assumptions, we can estimate the capillary quadrupolar potential between two microgels as a function of their center-to-center distance $d$ as 

\begin{equation}
  V(d)\approx-12\pi\gamma_{w,o}H_{1}H_{2}cos(2\varphi_{1}-2\varphi_{2})\dfrac{R^{4}}{d^{4}},
\end{equation}

where $j=1,2$ denote the two considered particles, $\gamma_{w,o}\approx55\,mN/m$ for a water/hexadecane interface, $H_{j}$ represent the amplitude of the interface deformation and $\varphi_{j}$ their phase, and $R$ is the radius of the particles at the interface \cite{Danov_JCIS_2005}. This expression is valid for $d>>2R$, but it is still in the same order of magnitude even in the limit $d\approx2R$, in the near field \cite{Near_field}.

Taking the example of 7CS1 microgels, we measure an experimental value of $d \approx1.8\,\mu m$ between particles in rectangular configurations, while $2R=1.2\mu m$ (see ESI). Assuming $T=25\,^\circ C$, and the orientation of the phase angles that gives the maximum attraction, we find that $H_{1,2}=[0.4,1.3]\,nm$ for $V(1.8\,\mu m)=[1,10]\,K_BT$.
These interface roughness values are compatible with the ones we measure from the outer part of the microgels via AFM, after deposition on a substrate, as reported here (see ESI) and in previous works \cite{Camerin_ACSNano_2019}. These considerations indicate that the formation of four-fold coordinated structures can emerge as a consequence of attractive capillary interactions with a quadrupolar leading order balanced by steric repulsion upon contact in the absence of external compression, i.e. for $\Pi\approx0\,mN/m$.   

\begin{figure}[h!]
\centering
  \includegraphics[width=\columnwidth]{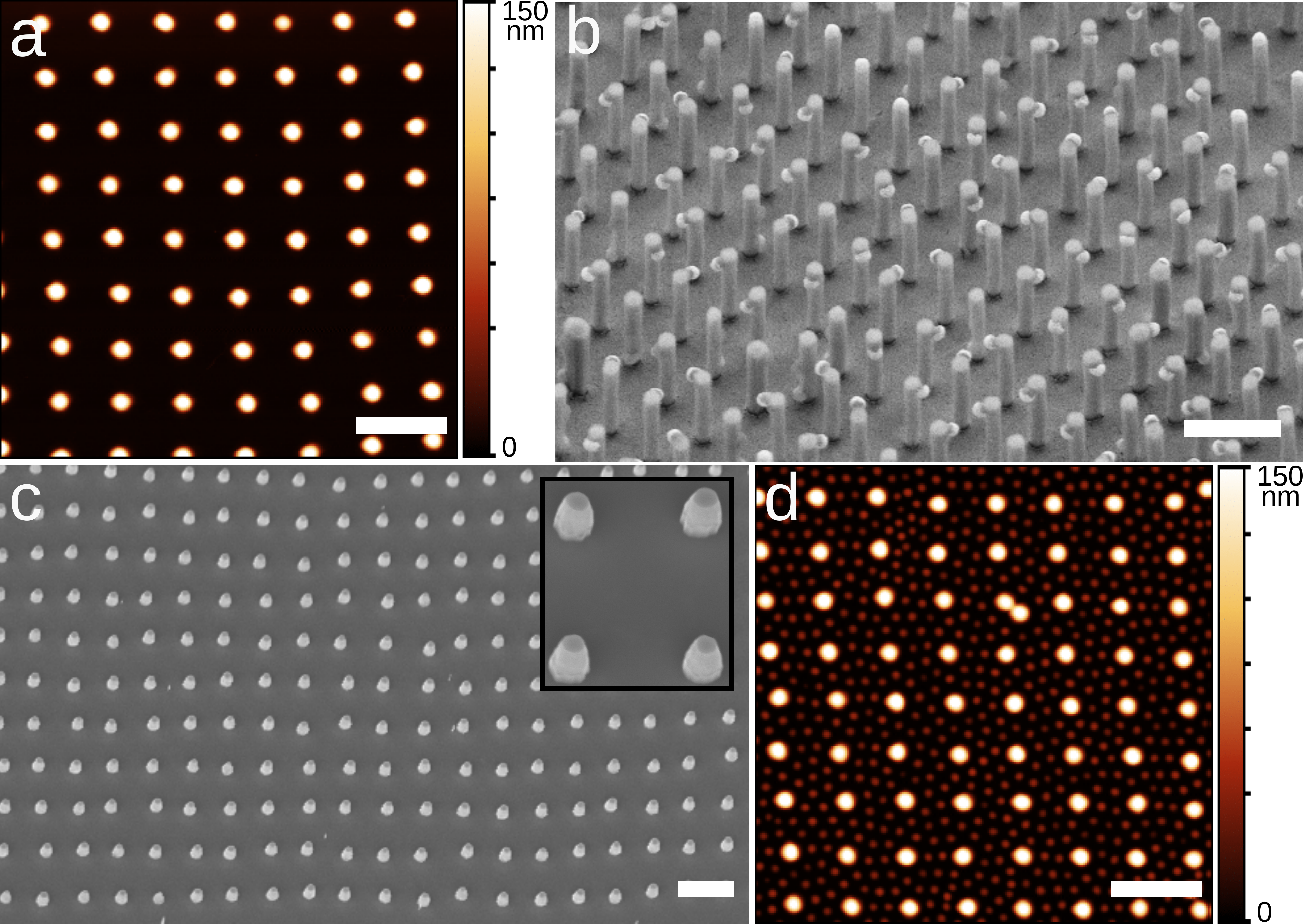}
  \caption{Square lattices for colloidal soft lithography applications. \textbf{(a)} AFM height image of a square lattice of C7S0 microgels deposited on a silicon substrate at $\Pi\approx0\,mN/m$. \textbf{(b)} Substrate prepared similar to the one in (a) and subjected to metal assisted chemical wet etching. \textbf{(c)} Square lattice of C3S0 microgels deposited on a silicon wafer and dry etched with deep reactive ion exchange. \textbf{(d)} Same substrate as in (a) subjected to a second deposition of smaller microgels with $\Pi\approx25\,mN/m$. All scale bars are $2\,\mu m$. The inset in (b) is $2\,\mu m$-wide.}
  \label{fig:3}
\end{figure}

Exploiting the spontaneous formation of non-hexagonal structures opens up new opportunities for soft colloidal lithography. As a demonstration, Fig. \ref{fig:3}a shows a square array of C7S0 microgels, which are subsequently used as a mask for the metal-assisted chemical "wet" etching, of vertically aligned silicon nanowires (VA-NWs) in a square lattice configuration (Fig. \ref{fig:3}b). To achieve this, a silicon substrate similar to the ones in Fig. \ref{fig:0} was prepared. In a next step, the microgels were swollen in photoresist, effectively turning into lithography masks \cite{Rey_NanoLett_2016} (see ESI). After the swelling, the substrate was sputter-coated by a $10\,nm$-thick gold layer and immersed in an etching solution containing $HF$ and $H_2O_2$ for 4 min. The photoresist-swollen microgels protect the underlying silicon wafer from the metal-assisted etching, leading to the formation of nanowires in correspondence of each microgel. The final structure was imaged via SEM, tilting the substrate by $30^\circ$. The VA-NWs are $\approx2.6\,\mu m$-long and the partially removed microgels in the washing process are still visible, and can be removed via further $O_2$ plasma treatment \cite{Fernandez-Rodriguez_Nanoscale_2018}. 
Furthermore, we tested the capability of such soft colloidal masks to be used in conventional deep reactive ion exchange, "dry" etching. To this end, we deposited C3S0 microgels in a square lattice configuration and swelled them in photoresist. Next, we performed the dry etching for 60s, obtaining the VA-NWs shown in the SEM image in Fig. \ref{fig:3}c. The pillars were $\approx640\,nm$-long and the microgels were etched away in the process. Although the dry etching could not produce VA-NWs as long as in the equivalent wet etching process, it leads to a very homogeneous patterning over the whole substrate, while wet etching is more heterogeneous and leads to a wider distribution of VA-NWs lengths. Ideally, one could improve the performance of soft colloidal masks for dry etching by improving the photoresist swelling process and combining it with a Bosch etching process where etching steps are alternated with the deposition of protective passivizing layers \cite{Bosch_process}. The use of microgels with different core-to-shell size ratios allows for the independent tuning of diameter and spacing of the VA-NWs.

Finally, Fig. \ref{fig:3}d, shows the possibility of realizing sequential depositions of microgels where the first layer has a square symmetry instead of a hexagonal one, as previously shown \cite{Fernandez-Rodriguez_Nanoscale_2018,Nature_2D}. First, a square lattice of C7S0 is produced at $\Pi\approx0\,mN/m$ (Fig. \ref{fig:3}a). Next, the same substrate is immersed in water and a second deposition of smaller P(NIPAM-co-MAA) microgels at $\Pi\approx25\,mN/m$ is performed to produce a 2D binary colloidal alloy (see ESI for further details). The combination of different symmetries for sequentially deposited microgel layers offers exciting opportunities for the future realization of complex structures.

Concluding, we show that interesting possibilities emerge from the interfacial self-assembly of soft colloids at close to zero surface pressures. This region of the compression curves is seemingly overlooked, as usually one finds 2D random gas and liquid phases for repulsion-dominated systems, or isolated aggregates if there are strong attractive forces. Here, by carefully exploring this region for soft microgels, we find that, at the right values of surface coverage, large-scale assemblies with four-fold symmetry emerge, which we hypothesize are caused by weak capillary attraction counterbalanced by stiff steric repulsion. These assemblies are found for a broad range of core-shell PNIPAM microgels, with different sizes and crosslinking densities, and for both water/oil and water/air interfaces. Their formation thus seems general and future work to prove if the same structures are found for microgels with different chemistries and architectures will be very interesting. We expect that these lattices with four-fold symmetries can not only be added to the extensive library of self-assembled 2D structures for colloidal lithography, but that they can also inspire further engineering of particle pair potentials at interfaces, e.g. through targeted particle-induced deformation of the interface as realized for larger objects \cite{Bae}. Detailed numerical studies of the interplay between particle architecture and interactions at the interface, paying close attention to near-field capillary forces, may finally direct the design of tailored soft colloids for the fabrication of new functional materials.

\section*{Conflicts of interest}
There are no conflicts to declare.

\section*{Author contributions}
Author contributions are defined based on the CRediT (Contributor Roles Taxonomy) and listed alphabetically. Conceptualization: MNA, MAFR, LI. Formal analysis: MNA, MAFR. Funding acquisition: LI. Investigation: MNA, MAFR. Methodology: MNA, MAFR, LI. Project administration: LI. Software: MAFR. Supervision: MAFR, LI. Validation: MNA, MAFR. Visualization: MNA, MAFR, LI. Writing original draft: MAFR, LI. Writing review and editing: MNA, MAFR, LI.

\section*{Acknowledgements}
All authors acknowledge Prof. Walter Richtering for providing the P(NIPAM-co-MAA) microgels. M.A.F.R. and L.I. acknowledge financial support from the Swiss National Science Foundation Grant PP00P2-172913/1. M.A.F.R. has received funding from the postdoctoral fellowships programme \textit{Beatriu de Pin{\'o}s}, funded by the Secretary of Universities and Research (Government of Catalonia) and by the Horizon 2020 programme of research and innovation of the European Union under the Marie Sklodowska-Curie grant agreement No 801370 (Grant 2018 BP 00029) and the Spanish \textit{Juan de la Cierva} Programme 2018 - Incorporation Grants (IJC2018-035946-I).



\balance


\bibliography{rsc} 
\bibliographystyle{rsc} 

\newpage
\noindent\LARGE{\textbf{Supplementary Information: Near-zero surface pressure assembly of rectangular lattices of microgels at fluid interfaces for colloidal lithography}} \\

\begin{center}
\large{Miguel Angel Fernandez-Rodriguez,\textit{$^{a,b}$$^{\ast}$} Maria-Nefeli Antonopoulou,\textit{$^{b,c}$} and Lucio Isa\textit{$^{b}$} } \\
\end{center}

\let\thefootnote\relax\footnotetext{\textit{$^{a}$}~\textit{Laboratory of Surface and Interface Physics, Biocolloids and Fluid Physics group, Department of Applied Physics, Faculty of Sciences, University of Granada, Campus de Fuentenueva s/n, ES 18071 Granada, Spain. E-mail: mafernandez@ugr.es}}
\footnotetext{\textit{$^{b}$}~\textit{Laboratory for Soft Materials and Interfaces, Department of Materials, Swiss Federal Institut of Technology Z\"urich, Vladimir-Prelog-Weg 1-5/10, 8093 Z\"urich, Switzerland.}}
\footnotetext{\textit{$^{c}$}~\textit{Current address: Polymeric Materials, Department of Materials, Swiss Federal Institut of Technology Z\"urich, Vladimir-Prelog-Weg 1-5/10, 8093 Z\"urich, Switzerland.}}

\renewcommand{\thefigure}{S\arabic{figure}}
\setcounter{figure}{0}    

\normalsize
\section*{Microgel synthesis}
We synthesized PNIPAM microgels by precipitation polymerization \cite{Camerin_ACSNano_2019}. We started from 180 mM of N-isopropyl\-acrylamide (NIPAM, TCI 98.0\%) monomer solutions in  MilliQ water, after purification and recrystallization in a hexane/toluene mixture. We then added the crosslinker N-N’-Methylenebisacryl-amide (BIS, Fluka 99.0\%) in different ratios, i.e.  at 3, 5, and 7 wt$\%$, respectively adjusting the volumes to have a constant total mass of monomers. The solution was kept at 80 $^\circ$C under $N_2$ atmosphere and the initiator potassium persulfate (KPS, Sigma-Aldrich 99.0\%) was added at 1.8 mM to initiate the reaction, keeping the temperature constant for 5 h. Next, the dispersion was cleaned three times by ultracentrifugation at 20000 rpm for 1 h. At the end of each ultracentrifugation cycle, the supernatant was replaced with MilliQ water, and the microgels were re-dispersed by 1 h of ultrasonication. These microgels present a core-shell architecture, with a more crosslinked core and a less crosslinked shell \cite{Camerin_ACSNano_2019}. The size of the particles can be increased, while keeping the same crosslinking ratio, by growing an extra shell. The reaction proceeds similarly to the one described above where the microgels are dispersed along the extra NiPAm and BIS monomers. The monomers were added in four steps by adding 1 mL of 1.2 mM KPS in MilliQ water and 1.25 mL of the crosslinker-monomer solution every 10 minutes. Afterwards, the synthesis and cleaning proceed as described before. We label our particles as CXSY, where X = 3,5,7 is the crosslinking mass ratio and Y=0 or 1 indicates the presence of an extra shell \cite{Nature_2D}. Additionally, we used a microgel with methacrylic acid co-monomer, P(NIPAM-co-MAA), synthesized as described in a previous publication \cite{Fernandez-Rodriguez_Nanoscale_2018} for the realisation of the sequential depositions. The sizes of all microgels in bulk MilliQ-water were measured by dynamic light scattering (DLS, Malvern Zetasizer) at 25 $^\circ$C. The diameter of isolated microgels deposited on silicon wafers (see next section for details on the deposition process) from the water/hexane interface were measured with atomic force microscopy (AFM, Brucker Icon Dimension) in tapping mode (300 kHz, $26\,mN/m$) at a rate of 1 Hz (see Table S1).

\begin{table}[H]
    \centering
    \begin{tabular}{|c|c|c|c|}\hline
    Microgel & $2R_h\,(nm)$ & $2R_{AFM}\,(nm)$ \\\hline
    C7S1    & $620\pm204$ & $1202\pm47$\\\hline
    C7S0    & $574\pm73$ & $786\pm30$\\\hline
    C5S1 & $620\pm204$ & $1066\pm71$\\\hline
    C5S0    & $597\pm127$ & $882\pm30$\\\hline
    C3S1 & $879\pm121$ & $1578\pm46$\\\hline
    C3S0    & $618\pm83$ & $1095\pm66$\\\hline
    P(NIPAM-co-MAA) & $213\pm10$ & $546\pm50$ \\\hline
    \end{tabular}
    \let\nobreakspace\relax\caption{Hydrodynamic diameter $2R_h$ of the microgels in bulk (DLS), and at the interface (from 25 isolated deposited microgels, by AFM).}
    \label{table:mgel_size}
\end{table}

\section*{Compression curve and depositions on silicon substrates at fixed $\Pi$}
\begin{figure}[H]
\centering
  \includegraphics[width=0.94\columnwidth]{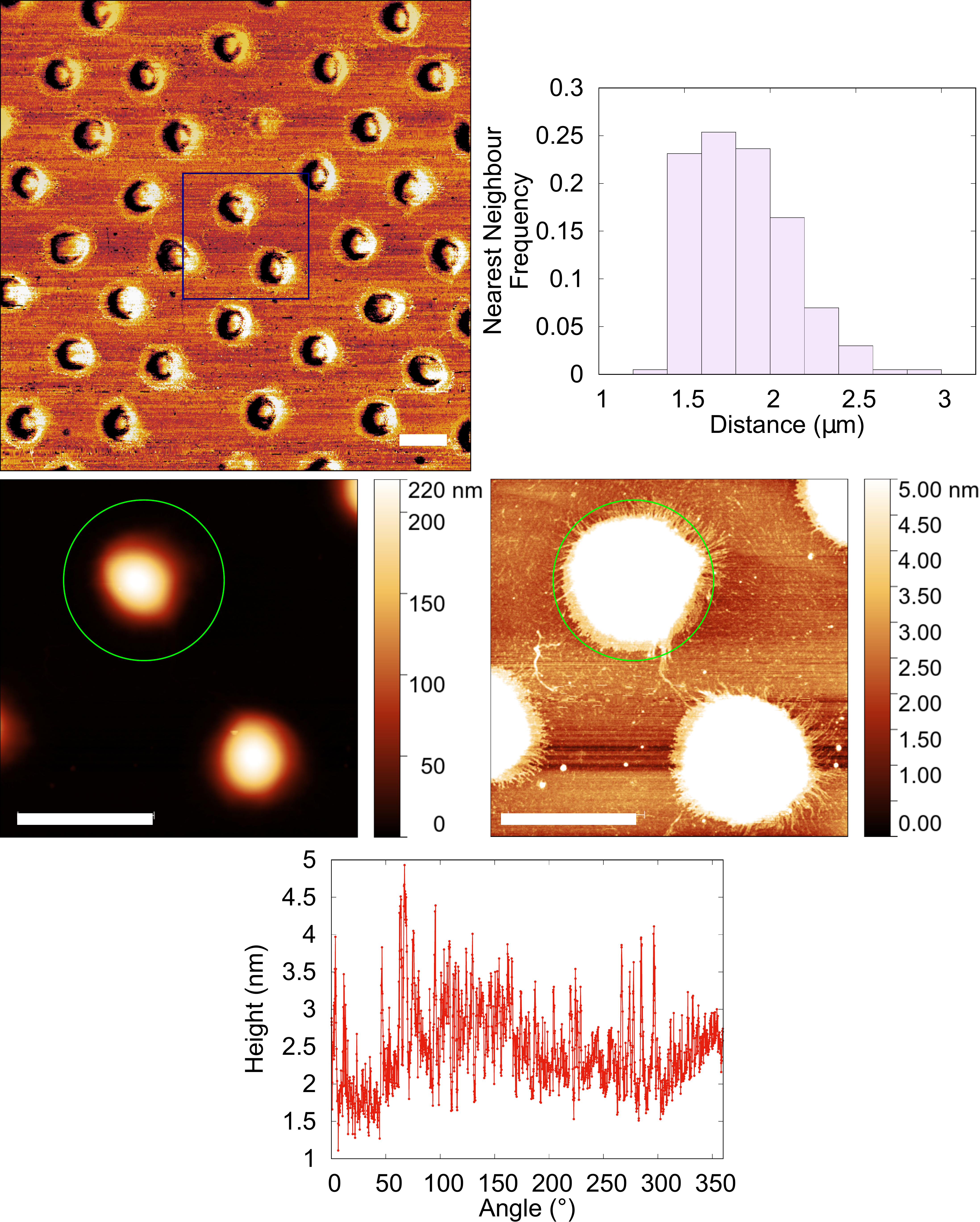}
  \let\nobreakspace\relax\caption{AFM phase image of square lattices obtained by depositing C7S1 microgels at $\Pi\approx0\,mN/m$. The histogram of nearest neighbours distribution shows distances clearly larger than the diameter reported in Table S1 and visible from the AFM images, with an average value of $(1.8\pm0.3)\,\mu m$. The AFM height images below are zoomed images, corresponding to the square marked in the image above. The height image on the left spans over the full height range while the one on the right is constrained to 5 nm-height to visualize the dangling polymers. A height profile is extracted at the green circle, which it is represented in the image below, with an average and standard deviation of $(2.5\pm0.6)\,nm$, and a maximum height difference of $3.8\,nm$. All scale bars are $1\,\mu m$.}
  \label{figS:AFM}
\end{figure}

Microgels were deposited on 2x2 cm$^2$ silicon substrates (Siltronix, $<100>$ ) as reported in previous works \cite{Rey_SoftMatter_2016,Scheidegger_PCCP_2017,Fernandez-Rodriguez_Nanoscale_2018,Nature_2D}. The silicon substrates (Siltronix, $<100>$ orientation, 100 mm single polished side, p-type, boron-doped, $3-6\,\Omega cm$) were cut into 2x2 $cm^2$ pieces with a diamond pen, cleaned in ultrasonic baths of toluene (Fluka Analytical, 99.7\%), isopropanol (IPA, Fisher Chemical, 99.97\%) and MilliQ water, and dried with N$_2$. The clean substrate was placed on the dipping arm of a liquid-liquid Langmuir-Blodgett trough (KSV5000, Biolin Scientific), at an orientation of 30$^\circ$ relative to the interface. We filled the trough with MilliQ water, keeping the substrate fully immersed, and cleaned the interface by closing the barriers and measuring the surface pressure $\Pi$ with a Wihelmy plate. Whenever $\Pi>0.4\,mN/m$, we vacuum cleaned the interface and opened the barriers, zeroed $\Pi$ and repeated the process again. Next, 100 mL of n-hexane (Sigma-Aldrich, HPLC grade 95\%, used as received) were added to create the water/hexane interface. Next, we raised the substrate until it barely crossed the water/hexane interface and zeroed the surface pressure $\Pi$. This position on the substrate is used a reference to estimate the value of $\Pi$ vs position. Next, we mixed 0.1 wt\% of the microgel suspensions in 4:1 suspension:IPA, where IPA acts as spreading agent at the interface. We then added the microgel dispersion at the interface with a Hamilton glass microsyringe (100 $\mu L$) and let the system stabilize for 10 minutes. Finally, for the compression curve $\Pi$ was gradually increased by compressing the interface with the barriers while the substrate was raised to deposit the monolayer at the same time. In the case of depositions with fixed $\Pi$, the position of the barrier is adjusted while the substrate was raised through the interface by the feedback of the Langmuir-Blodgett trough. The area per particle $A_p$ of the deposited monolayers for the compresion curve in Fig. 2 was obtained via optical microscopy (Zeiss Axio Imager Z1 Upright microscope) with a 63x objective and images taken with an AxioCam camera, processed with the open source software ImageJ. For the calculation of the $\Psi_4$ and $\Psi_6$ values a custom made python program using the freud package was used \cite{freud2020}. See Fig. \ref{figS:AFM} for an AFM characterisation of the square lattices. In this AFM image, it is worth noting that the standard deviation of a height profile on the periphery of a microgel is $\approx0.6\,nm$. This value is compatible with the amplitude of the capillary deformation $H_{1,2}=0.4\,nm$  estimated in the main text for a microgel at the water/oil interface, to explain a quadrupolar interaction of order $K_BT$. Thus, the roughness induced by the dangling polymers at the outermost part of the microgels at the interface might be enough to provide the quadrupolar capillary interaction that we observe.

\section*{Imaging the microgels in situ at interfaces}
\subsection*{Real-time imaging}
\begin{figure}[H]
\centering
  \includegraphics[width=0.8\columnwidth]{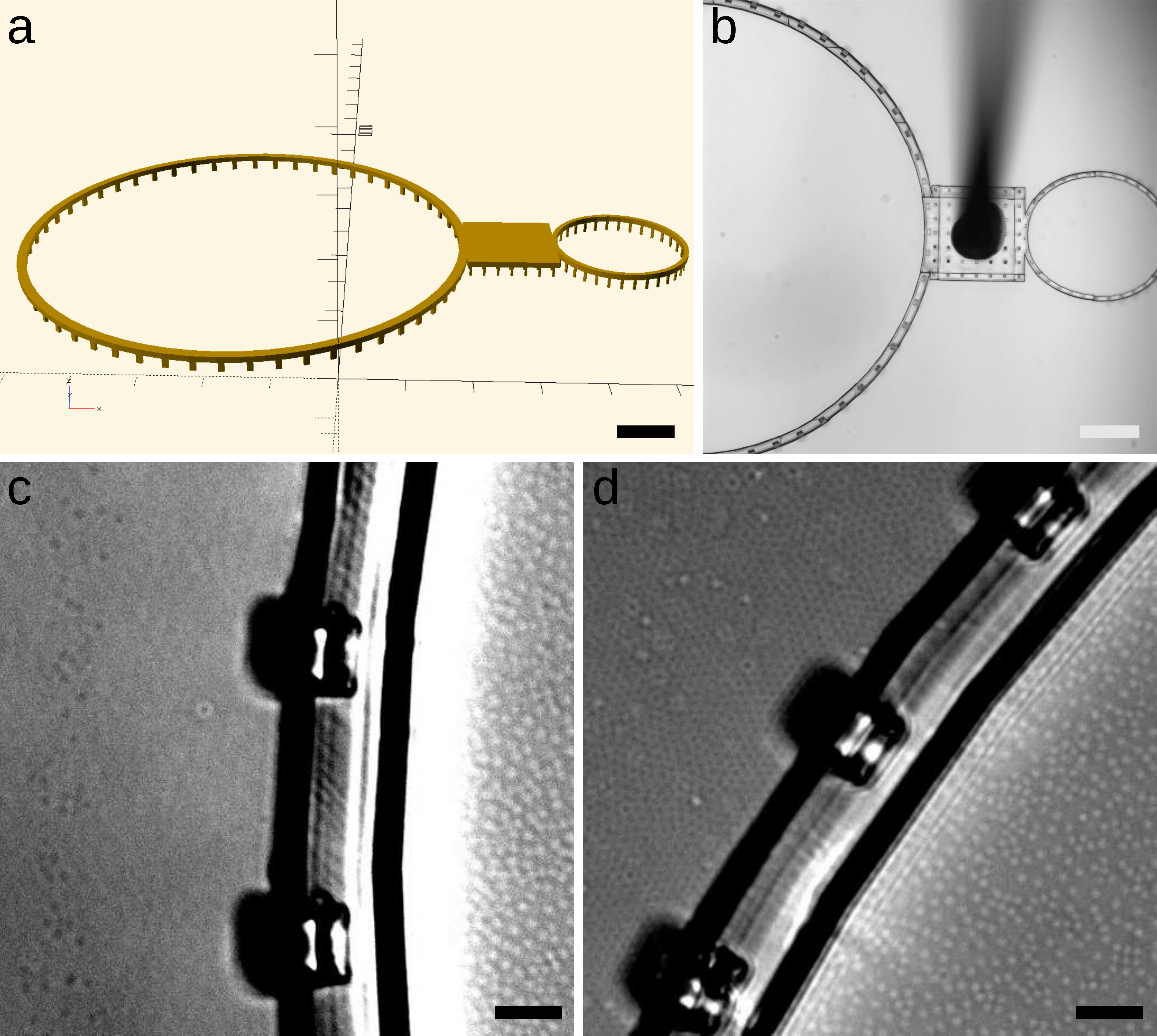}
  \let\nobreakspace\relax\caption{\textbf{(a)} Openscad model of the ring printed using the Nanoscribe GT2. The pillars are necessary to remove the device from the substrate where it is printed. The larger ring is used in the experiment. \textbf{(b)} Ring as printed, developed, cured over 2 days and connected to a $58\,\mu m$-diameter tungsten filament glued with araldite (the glue is visible as the black spot on the device). \textbf{(c-d)} Two different pictures during the same experiment with a C7S1 microgel-laden water/hexadecane interface imaged with a 40x dipping objective. In both pictures the right part is inside the ring. The left part shows a wide variation of $A_p$, ranging from sparse microgels (c) to a full monolayer (d) due to the convection flows present in the system. Scale bars are $100\,\mu m$ for (a-b) and $10\,\mu m$ for (c-d).}
  \label{figS:Nanoscribe}
\end{figure}

For the real time imaging we used two setups. One involves a custom-built liquid-liquid Langmuir trough (based on a KSV Nima medium trough) where the microgels are imaged with an upright microscope (Nikon LV-FM microscope equipped with a 40x dipping objective and a Coolsnap Dyno Mono camera - Movie S1, acquired at 10 fps). The ring that we used to restrict convection at the interface was 3D-printed with a Nanoscribe Professional GT device. The ring was designed in openscad (see Fig. \ref{figS:Nanoscribe}a), printed and developed, cured under UV light for 2 days, glued to a tungsten filament of $58\,\mu m$-diameter with araldite, micromanipulated with two micro-robots (miBOT, Imina Technologies, see Fig. \ref{figS:Nanoscribe}b) and placed on a C7S1 microgel-laden water/hexadecane interface (see Fig. \ref{figS:Nanoscribe}c-d). The second setup uses a conventional cell made of a big coverslip as a base, a smaller coverslip with a circular opening made by a laser cutter, and a metal ring on top. The three parts are glued together with UV glue. This enables to fill the well made between the two bottom coverslips with water, then the microgels are deposited and finally the cell is filled with hexadecane (Sigma Aldrich 99\%) and closed with another coverslip. Using a thin water well and a thin bottom glass enables to image the microgels at the interface with  with an inverted microscope (Zeiss Axio Observer Z1) using a 63x objective and acquiring images with a camera (sCMOS Andor Zyla) at 10 fps and 10 ms exposure time an inverted microscope using a 63x objective (Movie S2 and S3, recorded at 10 fps. Drift is substracted from the movies). No significant differences are expected between hexane and hexadecane due to similar interfacial tension and PNIPAM solubility.

\subsection*{Freeze-fracture shadow-casting cryo-SEM}
\begin{figure}[H]
\centering
  \includegraphics[width=\columnwidth]{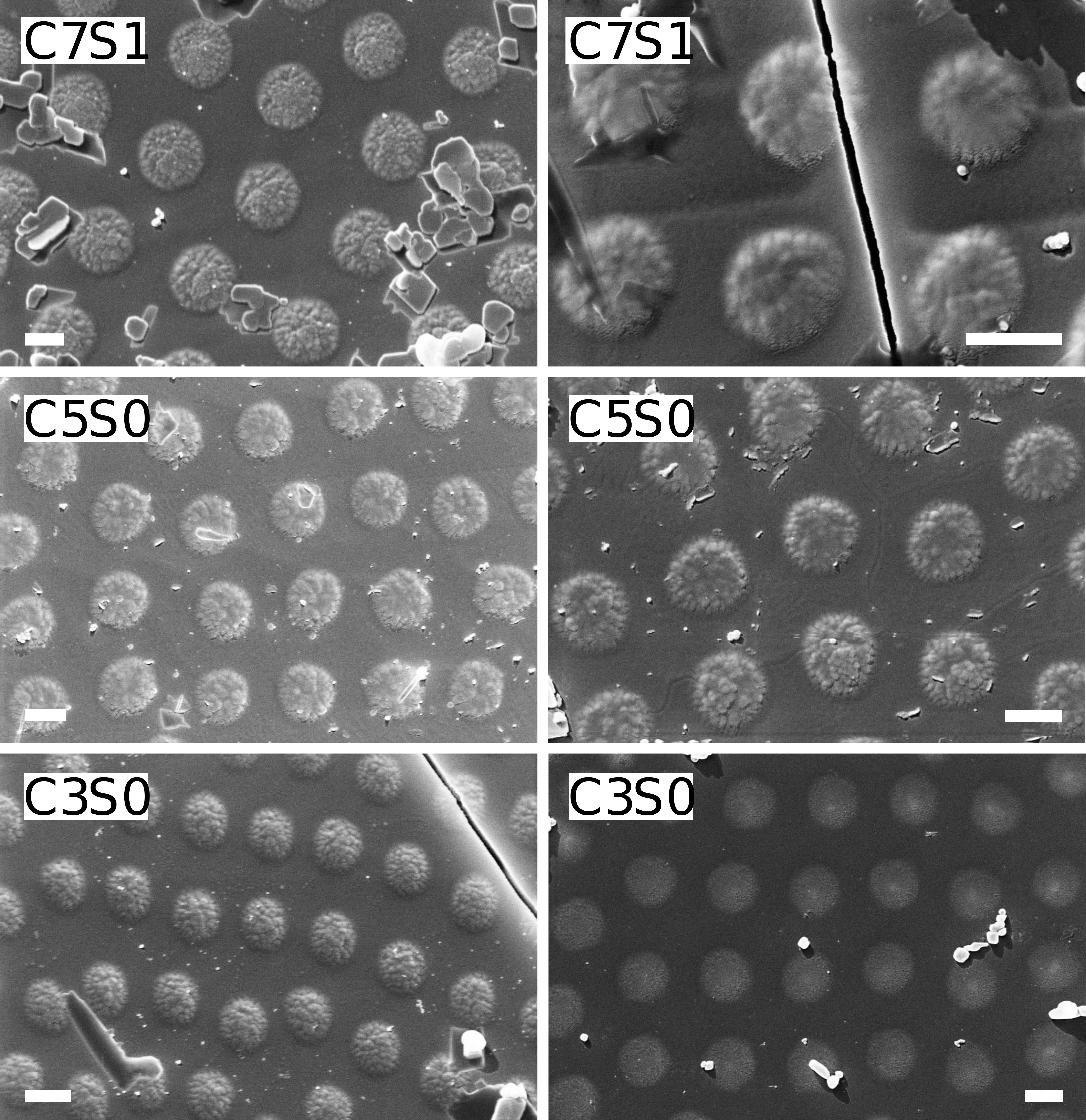}
  \let\nobreakspace\relax\caption{Cryo-SEM images of microgels at interfaces small domains with square lattices. Scale bars are $500\,nm$.}
  \label{figS:S_cryo}
\end{figure}

The freeze-fracture shadow-casting (FreSCa) cryo-SEM technique was used to image microgels at a water/decane (Sigma-Aldrich, 99\%) interface as described in a previous work \cite{Camerin_ACSNano_2019}. No significant differences are expected with hexane, as with hexadecane, due to similar interfacial tension and PNIPAM solubility. Microgel-laden interfaces were prepared by pipetting $0.5\,\mu L$ of the desired microgel dispersion at 0.1 wt\% into a $1\,mm^2$ copper holder and then covering it with $3\,\mu L$ of  decane purified in a basic alumina column. A second flat copper top enclosed the interface and the copper "sandwich" was vitrified with a propane jet freezer (Bal-Tec/Leica JFD 030) and fractured under high-vacuum and cryogenic conditions (Bal-Tec/Leica VCT010 and Bal-Tec/Leica BAF060). The system breaks at the weakest point, which is the water/decane interface, exposing the microgels protruding from the water phase. Next, we deposited 2 nm of tungsten forming a $30^\circ$-angle with the interface and imaged the sample in a cryo-SEM (Zeiss Leo 1530). As none of the microgels cast any shadow, this ensures the contact angle is $<30^\circ$. This confirms that they are hydrophilic and that the portion protruding into decane collapses due to poor solubility. In Fig. \ref{figS:S_cryo}, we show square lattices found for different microgels.

\subsection*{Curing the interface to characterize it via AFM and SEM}
\begin{figure}[H]
\centering
  \includegraphics[width=0.8\columnwidth]{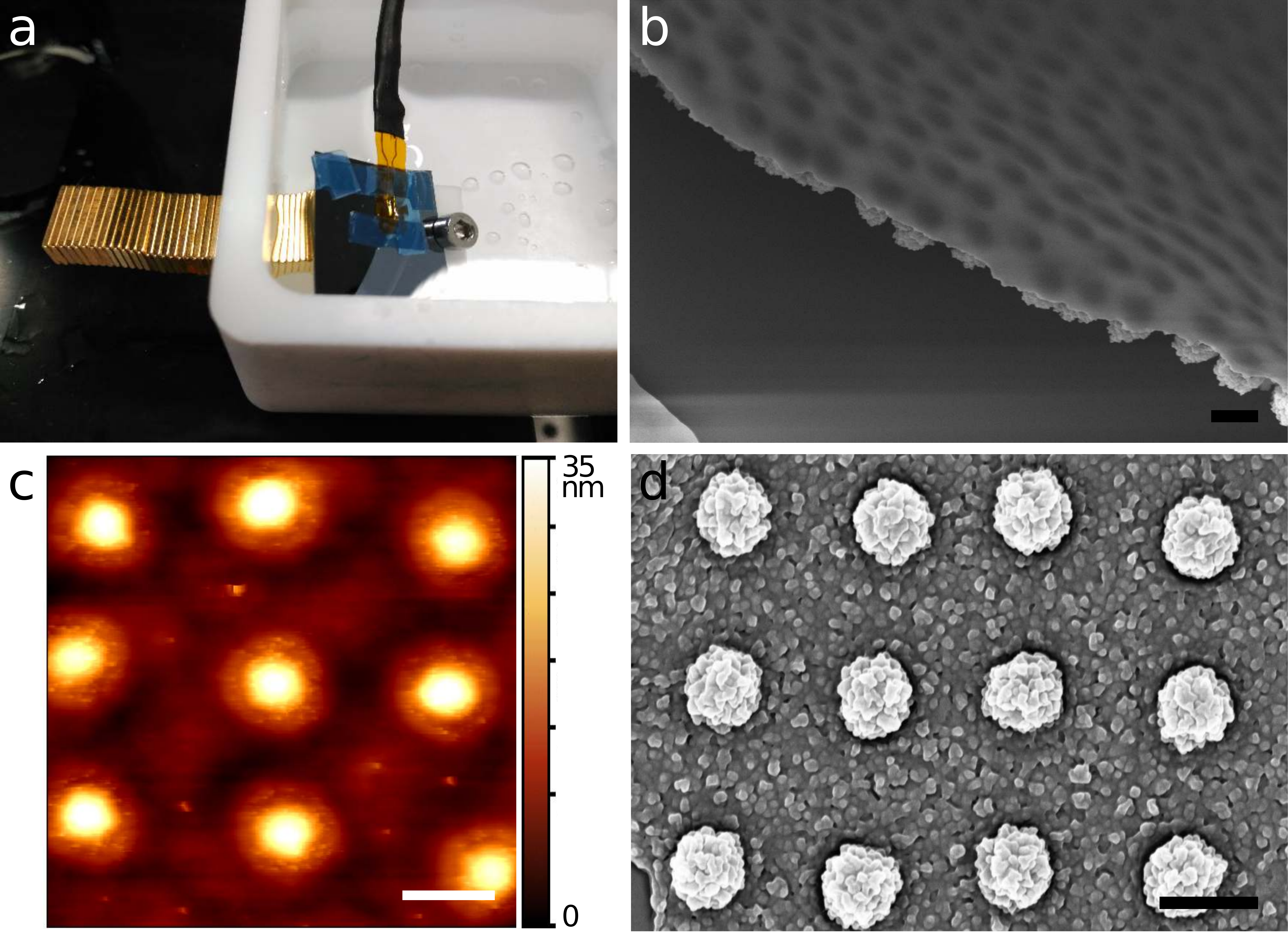}
  \let\nobreakspace\relax\caption{C7S1 microgel-laden water/air interface cured with cyanoacrylate. \textbf{(a)} Teflon beaker with an immersed $2x2\,cm^2$ silicon substrate, tilted by $30^\circ$ relative to the interface and movable using the yellow magnets. The water temperature was measured with a thermocouple. \textbf{(b)} SEM image of a ruptured piece of the cured surface affording a lateral view of the film and the part of the microgels protruding into the water phase below the interface. \textbf{(c)} AFM height image of a cured microgel-laden water/air interface after exposure to cyanoacrylate \textbf{(d)} SEM image of an upside-down fragment of the microgel-cyanoacrylate film, showing the portion of the microgels that were immersed in water, with the microgels in a square lattice configuration. Scale bars $1\,\mu m$.}
  \label{figS:S_superglue1}
\end{figure}

A C7S1 microgel-laden water/air interface created at $\Pi\sim0$ in a teflon beaker (see Fig. \ref{figS:S_superglue1}a) was cured by exposing it to cyanoacrylate vapour overnight \cite{Weiss}. We deposit the cured interface onto a $2x2\,cm^2$ silicon substrate that was immersed in the water phase previous to the deposition of the microgels at the interface. The substrate is tilted $30^\circ$ with the interface with a teflon holder and it could be moved using the yellow magnets in the picture. After exposure to the cyanoacrylate vapour, the substrate was raised and the cured interface deposited on the silicon wafer. After transfer, the film ruptured in several locations, enabling the observation of the particles from the side (see Fig. \ref{figS:S_superglue1}b) and from below (see Fig. \ref{figS:S_superglue1}c-d, showing  the portion of the microgels that was originally immersed in water). We observed square lattices in these cured surfaces with core-core distances compatibles with the histogram in Fig. \ref{figS:AFM}. 

\newpage
\subsection*{Soft colloidal lithography}
\begin{figure}[H]
\centering
  \includegraphics[width=\columnwidth]{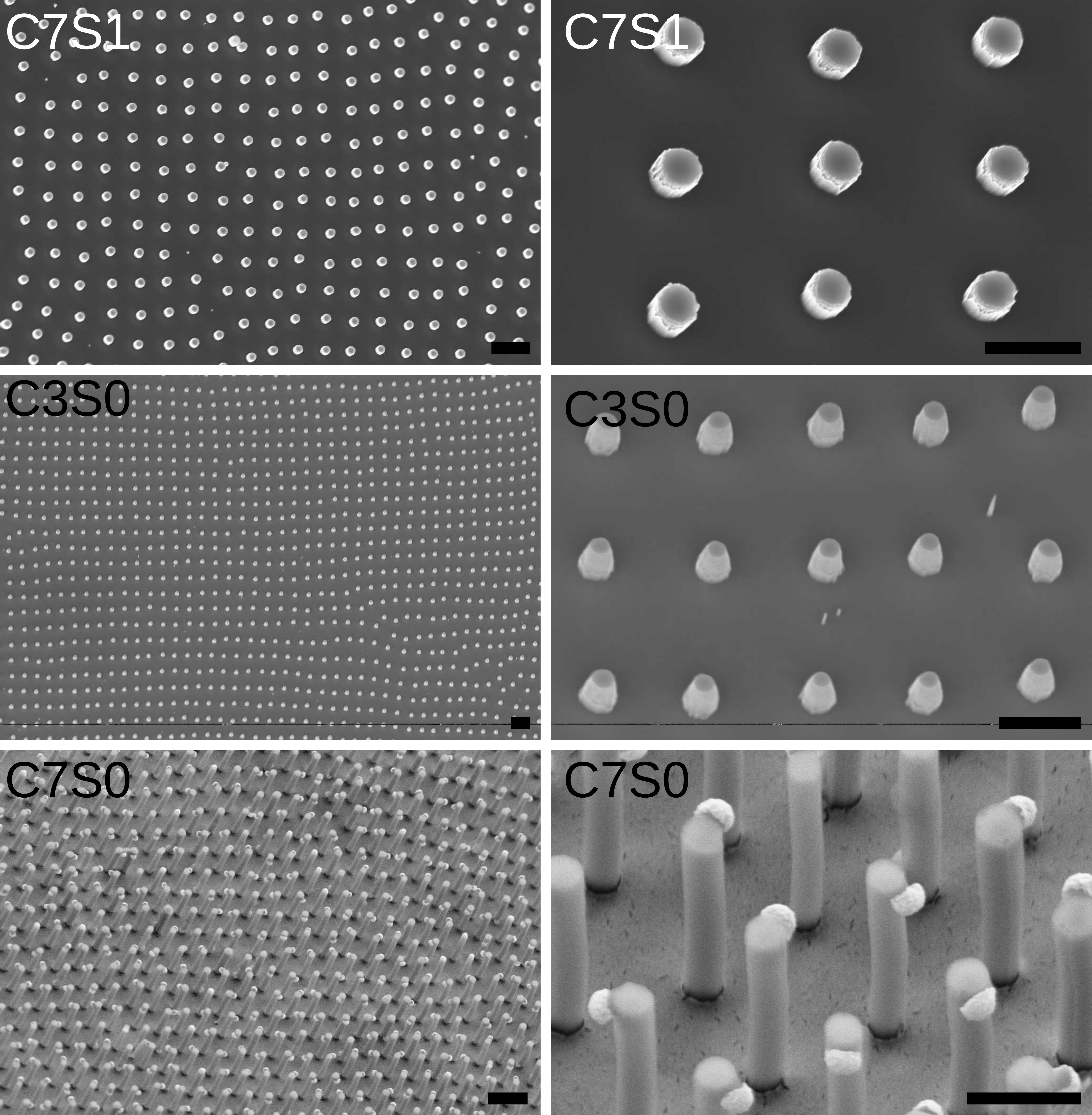}
  \let\nobreakspace\relax\caption{SEM images of vertically aligned nanowires obtained by soft colloidal lithography of square lattices deposited on silicon substrates. The substrates are tilted by $30^\circ$ to quantify the height of the nanowires (right column). The substrate with C7S1 microgels was subjected to 45 s of dry etching ($400\,nm$ pillars), the one with C3S0 to 30 s of dry etching ($640\,nm$ pillars), and the one with C7S0 to 4 min of wet etching ($2.6\,\mu m$ pillars).  Scale bars are $2\,\mu m$ for the panels in the left and $1\,\mu m$ for the panels in the right.}
  \label{figS:S_etching}
\end{figure}

Once the microgel monolayer is deposited on a silicon wafer, the microgels are very flat, as can be seen by Fig. 4a, where the maximum height of the microgels is 150 nm. This is often not enough to act to provide a sufficient masking for etching processes. We increased their thickness by exposing them to photoresist (AZ1518, MicroChemicals GmbH) \cite{Rey_NanoLett_2016} for 30 min, baking the substrate at $100\,^\circ C$ for 1 min and rinsing the excess of photoresist with acetone. This mechanism relies on the ability of the microgels to be swollen by the photoresist  dissolved in ethanol, and after baking, on the different rate of removal of the photoresist by acetone, being slower for the photoresist inside the microgel. The final result is that the microgels increase their thickness. For example, for C7S0 the height increases from the $150\,nm$ reported in Fig. 4a to $\sim500\,nm$, which it is comparable to the height provided by conventional spin coating of AZ1505 photoresist. Therefore, this intermediate step of swelling enables their use as etching masks in both metal-assisted chemical wet etching (MACE), and in deep reactive ion exchange dry etching (DRIE). For MACE, after swelling in photoresist, a $10\,nm$ layer of gold is sputter coated (Safematic CCU-010). Next, the substrate is submerged in the wet etching solution containing 3:3:2:2 solution of MilliQ water:ethanol:$HF$:$H_2O_2$ (ethanol from Fluka Analytical, 99.8\%; HF from Sigma Aldrich, 48\%; $H_2O_2$ from Merck 30\%). Next, the substrate is rinsed in 1:4:40 of $I_2$:$KI$:MilliQ water ($I_2$ and $KI$ from Sigma Aldrich, 99 \%) to remove the gold layer, and subsequent rinsing in ethanol and acetone (Sigma Aldrich, 99 \%). Next, for big aspect ratio-pillars, a $CO_2$ supercritical drier (Autosamdri-931, Tousimis) was used to avoid the collapse of the pillars. In the case of DRIE, the microgels swollen in photoresist are used directly as the masks for etching. Contrary to MACE, DRIE is harsh with the mask and removes it overtime. In 30-45s the microgels are etched away. This constrains the maximum height of the pillars but also opens new possibilities on engineering the shape of the pillars as can be seen in Fig. \ref{figS:S_etching}, as once the microgel is etched away, the pillars start being etched into a tapered shape (see the pillars obtained with C3S0 microgels). On the other hand, DRIE produces a very homogeneous etching over the whole substrate, which it is a challenge for MACE.

\end{document}